\documentclass[11pt,twoside,dvips,letterpaper]{paper}

\usepackage{graphicx}
\usepackage{amsmath}
\usepackage{amssymb}
\usepackage{bm}
\usepackage[usenames,dvipsnames]{color}
\usepackage{mathrsfs}
\usepackage[colorlinks,dvips]{hyperref}
\usepackage[hyphenbreaks]{breakurl}
\usepackage[margin=1in]{geometry}
\usepackage[compact]{titlesec}
\usepackage{mdwlist}
\usepackage{subcaption}

\newcommand{\bv}[1]{{\mbox{\boldmath$\mathbf{#1}$}}}

\newcommand{\rmean}[1]{{\left\langle{#1}\right\rangle}}
\newcommand{\irmean}[1]{{\langle{#1}\rangle}}

\newcommand{\rv}[1]{\rmean{#1^2}}

\newcommand{\Eqr}[1]{(\ref{#1})}
\newcommand{\Eqre}[1]{Eq.~(\ref{#1})}

\newcommand{\Eqres}[1]{Eqs.~(\ref{#1})}

\newcommand{\laur}{LA-UR 13-21573}

\newcommand{\ti}{A stochastic diffusion process for Lochner's generalized
                 Dirichlet distribution}
\newcommand{\aufirst}{J.\ Bakosi}
\newcommand{\ausecond}{J.R.\ Ristorcelli}
\newcommand{\au}{\aufirst and \ausecond}
\newcommand{\kw}{Fokker-Planck equation;
                 Stochastic diffusion;
                 Generalized Dirichlet distribution}
\hypersetup{citecolor=blue,
            linkcolor=blue,
            urlcolor=blue,
            pdftitle=\ti,
            pdfauthor=\au,
            pdfkeywords=\kw}
\usepackage{hypernat}

\begin{document}

\title{\ti\\[0.2cm]\small\texttt{\laur}\\[0.2cm]\texttt{Accepted for publication in
Journal of Mathematical Physics, September 12, 2013}\\[-0.3cm]}
\author{\normalsize\aufirst, \normalsize\ausecond\\[-0.1cm]
        \texttt{\normalsize\{jbakosi,jrrj\}@lanl.gov} \\[-0.1cm]
        \normalsize Los Alamos National Laboratory, Los Alamos, NM 87545,
        USA\\[-0.3cm]}

\maketitle

\begin{abstract}
The method of potential solutions of Fokker-Planck equations is used to develop
a transport equation for the joint probability of $N$ stochastic variables with
Lochner's generalized Dirichlet distribution \cite{Lochner_75} as its asymptotic
solution. Individual samples of a discrete ensemble, obtained from the system of
stochastic differential equations, equivalent to the Fokker-Planck equation
developed here, satisfy a unit-sum constraint at all times and ensure a bounded
sample space, similarly to the process developed in \cite{Bakosi_dir} for the
Dirichlet distribution. Consequently, the generalized Dirichlet diffusion
process may be used to represent realizations of a fluctuating ensemble of $N$
variables subject to a conservation principle. Compared to the Dirichlet
distribution and process, the additional parameters of the generalized Dirichlet
distribution allow a more general class of physical processes to be modeled with
a more general covariance matrix.
\end{abstract}

\section{Introduction}
\label{sec:introduction}
We develop a Fokker-Planck equation whose statistically stationary solution
(i.e.\ invariant) is Lochner's generalized Dirichlet
distribution \cite{Lochner_75,Connor_69,Wong_98}.

The (standard) Dirichlet distribution \cite{Johnson_60,Mosimann_62,Kotz_00} has
been used to represent a set of non-negative fluctuating variables subject to a
unit-sum requirement in a variety of fields, including evolutionary theory
\cite{Pearson_1896}, Bayesian statistics \cite{Paulino_95}, geology
\cite{Chayes_62,Martin_65}, forensics \cite{Lange_95}, econometrics
\cite{Gourieroux_06}, turbulent combustion \cite{Girimaji_91}, and population
biology \cite{Steinrucken_2013}. Following the method of potential solutions,
applied in \cite{Bakosi_dir}, we derive a system of coupled stochastic
differential equations (SDE) whose (Wiener-process) diffusion terms are
nonlinearly coupled and whose invariant is Lochner's generalized Dirichlet
distribution.

The standard Dirichlet distribution can only represent non-positive covariances
\cite{Mosimann_62}, which limits its application to a specific class of
processes. The stochastic system whose invariant is the generalized Dirichlet
distribution allows for a more general class of physical processes with a more
general covariance matrix. The process may be stationary or non-stationary, not
limited to non-positive covariances, and satisfies the unit-sum requirement at
all times, necessary for variables that obey a conservation principle.

\section{Preview of results}
The generalized Dirichlet distribution for a set of scalars $0 \le Y_i$,
$i=1,\dots,K$, $\sum_{i=1}^KY_i\le1$, and parameters, $\alpha_i>0$,
$\beta_i>0$, as given by Lochner \cite{Lochner_75} reads
\begin{equation}
\mathscr{G}(\bv{Y},\bv{\alpha},\bv{\beta}) =
\prod_{i=1}^K\frac{\Gamma(\alpha_i+\beta_i)}{\Gamma(\alpha_i)\Gamma(\beta_i)}
Y_i^{\alpha_i-1} \mathcal{Y}_i^{\gamma_i} \qquad \mathrm{with} \qquad
\mathcal{Y}_i = 1-\sum_{k=1}^i Y_k,
\label{eq:GD}
\end{equation}
and $\gamma_i=\beta_i-\alpha_{i+1}-\beta_{i+1}$ for $i=1,\dots,K-1$, and
$\gamma_K=\beta_K-1$. Here $\Gamma(\cdot)$ denotes the gamma function. We derive
the stochastic diffusion process, governing the scalars, $Y_i$,
\begin{equation}
\begin{split}
\mathrm{d}Y_i(t) = \frac{\mathcal{U}_i}{2}\left\{ b_i\Big[S_i \mathcal{Y}_K -
(1-S_i)Y_i\Big] + Y_i\mathcal{Y}_K
\sum_{j=i}^{K-1}\frac{c_{ij}}{\mathcal{Y}_j}\right\}\mathrm{d}t +
\sqrt{\kappa_i Y_i \mathcal{Y}_K \mathcal{U}_i}\mathrm{d}W_i(t), \\
\qquad i=1,\dots,K,
\end{split}
\label{eq:iSDE}
\end{equation}
where $\mathrm{d}W_i(t)$ is an isotropic vector-valued Wiener process with
independent increments \cite{Gardiner_09} and $\mathcal{U}_i = \prod_{j=1}^{K-i}
\mathcal{Y}_{K-j}^{-1}$. We show that the statistically stationary solution of
the coupled system of nonlinear stochastic differential equations in
\Eqr{eq:iSDE} is the generalized Dirichlet distribution, \Eqre{eq:GD}, provided
the coefficients, $b_i\!>\!0$, $\kappa_i\!>\!0$, $0\!<\!S_i\!<\!1$, and $c_{ij}$, with
$c_{ij}\!=\!0$ for $i\!>\!j$, $i,j\!=\!1,\dots,K\!-\!1$, satisfy
\begin{align}
\alpha_i & = \frac{b_i}{\kappa_i}S_i, \qquad i=1,\dots,K,\\
1-\gamma_i & = \frac{c_{1i}}{\kappa_1} = \dots = \frac{c_{ii}}{\kappa_i},
\qquad i=1,\dots,K-1,\\
1+\gamma_K & = \frac{b_1}{\kappa_1}(1-S_1) = \dots =
\frac{b_K}{\kappa_K}(1-S_K).
\end{align}
The restriction on the coefficients ensure reflection towards the interior of
the sample space, which together with the specification $Y_N=\mathcal{Y}_K$
ensures
\begin{equation}
\sum_{i=1}^NY_i=1. \label{eq:sum}
\end{equation}
Indeed, if for example $Y_1\!=\!0$, the diffusion in \Eqre{eq:iSDE} is zero and
the drift is strictly positive, while if $Y_1\!=\!1$, the diffusion is zero (as
$\mathcal{Y}_K\mathcal{U}_1\!\rightarrow\!0$) and the drift is strictly
negative.

\section{Development of the diffusion process}
\label{sec:sde}
The diffusion process, \Eqre{eq:iSDE}, is developed by the method of potential
solutions. The steps below closely follow the methodology introduced in
\cite{Bakosi_dir}, used to derive a diffusion process for the Dirichlet
distribution.

We start from the It\^o diffusion process \cite{Gardiner_09} for the stochastic
vector, $Y_i$,
\begin{equation}
\mathrm{d}Y_i(t) = a_i(\bv{Y})\mathrm{d}t + \sum_{j=1}^K
b_{ij}(\bv{Y})\mathrm{d}W_j(t),
\qquad\quad i = 1,\dots,K,\label{eq:Ito}
\end{equation}
with drift, $a_i(\bv{Y})$, diffusion, $b_{ij}(\bv{Y})$, and the isotropic
vector-valued Wiener process, $\mathrm{d}W_j(t)$. Using standard methods given
in \cite{Gardiner_09} the equivalent Fokker-Planck equation governing the joint
probability, $\mathscr{F}(\bv{Y},t)$, derived from \Eqre{eq:Ito}, is
\begin{equation}
\frac{\partial\mathscr{F}}{\partial t} = - \sum_{i=1}^K \frac{\partial}{\partial
Y_i}\big[a_i(\bv{Y})\mathscr{F}\big] + \frac{1}{2}\sum_{i=1}^K\sum_{j=1}^K
\frac{\partial^2}{\partial Y_i \partial
Y_j}\big[B_{ij}(\bv{Y})\mathscr{F}\big], \qquad\quad
B_{ij} = \sum_{k=1}^K b_{ik} b_{kj}.
\label{eq:FP}
\end{equation}
As the drift and diffusion coefficients are time-homogeneous,
$a_i(\bv{Y},t)\!=\!a_i(\bv{Y})$ and $B_{ij}(\bv{Y},t)\!=\!B_{ij}(\bv{Y})$,
\Eqre{eq:Ito} is a statistically stationary process and the solution of
\Eqre{eq:FP} converges to a stationary distribution \cite{Gardiner_09}, Sec.\
6.2.2. Our task is to specify the functional forms of $a_i(\bv{Y})$ and
$b_{ij}(\bv{Y})$ so that the stationary solution of \Eqre{eq:FP} is
$\mathscr{G}(\bv{Y})$, defined by \Eqre{eq:GD}.

A potential solution of \Eqre{eq:FP} exists if
\begin{equation}
\frac{\partial\ln\mathscr{F}}{\partial Y_j} = \sum_{i=1}^K B_{ij}^{-1}\left(2a_i
- \sum_{k=1}^K \frac{\partial B_{ik}}{\partial Y_k}\right) \equiv
-\frac{\partial\phi}{\partial Y_j}, \qquad\quad j = 1,\dots,K,
\label{eq:solution}
\end{equation}
is satisfied, \cite{Gardiner_09} Sec.\ 6.2.2. Since the left hand side of
\Eqre{eq:solution} is a gradient, the expression on the right must also be a
gradient and can therefore be obtained from a scalar potential denoted by
$\phi(\bv{Y})$. This puts a constraint on the possible choices of $a_i$ and
$B_{ij}$ and on the potential, as $\phi,_{ij}=\phi,_{ji}$ must also be
satisfied. The potential solution is
\begin{equation}
\mathscr{F}(\bv{Y}) = \exp[-\phi(\bv{Y})].\label{eq:phi}
\end{equation}
Now functional forms of $a_i(\bv{Y})$ and $B_{ij}(\bv{Y})$ that satisfy
\Eqre{eq:solution}, with $\mathscr{F}(\bv{Y}) \equiv \mathscr{G}(\bv{Y})$ are
sought. The mathematical constraints on the specification of $a_i$ and $B_{ij}$
are as follows:
\begin{enumerate*}
\item $B_{ij}$ must be symmetric positive semi-definite. This is to
ensure that
\begin{itemize*}
\item the square-root of $B_{ij}$ (e.g.\ the Cholesky-decomposition, $b_{ij}$)
exists, required by the correspondence of the stochastic equation \Eqr{eq:Ito}
and the Fokker-Planck equation \Eqr{eq:FP},
\item \Eqre{eq:Ito} represents a diffusion, and
\item $\det(B_{ij})\ne0$, required by the existence of the inverse in
\Eqre{eq:solution}.
\end{itemize*}
\item For a potential solution to exist \Eqre{eq:solution} must be satisfied.
\end{enumerate*}
With $\mathscr{F}(\bv{Y}) \equiv \mathscr{G}(\bv{Y})$ \Eqre{eq:phi} shows that
the scalar potential must be
\begin{equation}
-\phi(\bv{Y}) = \sum_{i=1}^K(\alpha_i-1)\ln Y_i + \sum_{i=1}^K\gamma_i\ln
\mathcal{Y}_i.\label{eq:gphi}
\end{equation}
It is straightforward to verify that the specifications
{\allowdisplaybreaks
\begin{align}
a_i(\bv{Y}) & = \frac{\mathcal{U}_i}{2}\left\{ b_i\Big[S_i \mathcal{Y}_K -
(1-S_i)Y_i\Big] + Y_i\mathcal{Y}_K \sum_{j=i}^{K-1}
\frac{c_{ij}}{\mathcal{Y}_j}\right\}, \label{eq:ga}\\
B_{ij}(\bv{Y}) & = \left\{ \begin{array}{lr} \kappa_i Y_i \mathcal{Y}_K
\mathcal{U}_i & \quad \mathrm{for} \quad i = j,\\\noalign{\smallskip} 0 & \quad
\mathrm{for} \quad i \ne j, \end{array} \right.\label{eq:gB}
\end{align}}%
satisfy the above mathematical constraints, 1.\ and 2. Here
$\mathcal{U}_i\!=\!\prod_{j=1}^{K-i} \mathcal{Y}_{K-j}^{-1}$, where an empty
product is assumed to be unity, while an empty sum is zero. In addition to the
coefficients $b_i\!>\!0$, $\kappa_i\!>\!0$, and $0\!<\!S_i\!<\!1$, governing the
Dirichlet diffusion process \cite{Bakosi_dir}, the drift now has the additional
(not all independent) ones, denoted by $c_{ij}$, with $c_{ij}=0$ for $i>j$,
$i,j=1,\dots,K-1$.

Substituting Eqs.\ (\ref{eq:gphi}--\ref{eq:gB}) into \Eqre{eq:solution} yields
a system with the same functions on both sides with different coefficients,
yielding the correspondence between the parameters of the generalized Dirichlet
distribution, \Eqre{eq:GD}, and the Fokker-Planck equation \Eqr{eq:FP} with
Eqs.\ (\ref{eq:ga}--\ref{eq:gB}) as
\begin{align}
\alpha_i & = \frac{b_i}{\kappa_i}S_i, \qquad i=1,\dots,K,\label{eq:galphai}\\
1-\gamma_i & = \frac{c_{1i}}{\kappa_1} = \dots = \frac{c_{ii}}{\kappa_i},
\qquad i=1,\dots,K-1,\label{eq:gammai}\\
1+\gamma_K & = \frac{b_1}{\kappa_1}(1-S_1) = \dots =
\frac{b_K}{\kappa_K}(1-S_K).\label{eq:gammaN}
\end{align}
The above result is arrived at inductively based on the special case of $K=3$ in
Appendix A. If Eqs.\ (\ref{eq:galphai}--\ref{eq:gammaN}) hold, the stationary
solution of the Fokker-Planck equation \Eqr{eq:FP} with drift \Eqr{eq:ga} and
diffusion \Eqr{eq:gB} is the generalized Dirichlet distribution, \Eqre{eq:GD}.
The same methodology was applied to the Dirichlet case in \cite{Bakosi_dir}.
Eqs.\ (\ref{eq:galphai}--\ref{eq:gammaN}) specify the correspondence between the
coefficients of the stochastic system \Eqr{eq:Ito} with drift \Eqr{eq:ga} and
diffusion \Eqr{eq:gB} and the generalized Dirichlet distribution, \Eqre{eq:GD}.
With $\gamma_i=\beta_i-\alpha_{i+1}-\beta_{i+1}$, $i=1,\dots,K-1$, and
$\gamma_K=\beta_K-1$, the correspondence between ($\alpha_i,\beta_i$) and
($b_i,S_i,\kappa_i,c_{ij}$) is also complete. Note that Eqs.\
(\ref{eq:ga}--\ref{eq:gB}) are one possible way of specifying drift and
diffusion to arrive at a generalized Dirichlet distribution; other functional
forms may be possible. It is straightforward to verify, that setting
$c_{1i}/\kappa_i=\dots=c_{ii}/\kappa_i=1$ for $i=1,\dots,K\!-\!1$, i.e.,
$\gamma_1=\dots=\gamma_{K-1}=0$, in \Eqres{eq:ga} and \Eqr{eq:gB} yields the
same system in \Eqre{eq:solution} as with $a_i$ and $B_{ij}$ specified for the
(standard) Dirichlet distribution, see Appendix A for $K=3$. The shape of the
generalized Dirichlet distribution, \Eqre{eq:GD}, is determined by the $2K$
coefficients, $\alpha_i$, $\beta_i$. Eqs.\ (\ref{eq:galphai}--\ref{eq:gammaN})
show that in the stochastic system, different combinations of $b_i$, $S_i$,
$\kappa_i$, and $c_{ij}$ may yield the same $\alpha_i$, $\beta_i$ and that not
all of $b_i$, $S_i$, $\kappa_i$, and $c_{ij}$ may be chosen independently to
make the invariant generalized Dirichlet. In other words, a unique set of SDE
coefficients always corresponds to a unique set of distribution parameters, but
the converse is not true: a set of distribution parameters do not uniquely
determine all the SDE coefficients, for a given specific asymptotic generalized
Dirichlet distribution.

\section{Properties of Dirichlet distributions}
It is useful to show how the generalized Dirichlet distribution, \Eqre{eq:GD},
reduces to standard Dirichlet, and their univariate case, the beta distribution.

\subsection{Density functions}
Setting $\gamma_1=\dots=\gamma_{K-1}=0$ in \Eqre{eq:GD} yields the (standard)
Dirichlet distribution
\begin{equation}
\mathscr{D}(\bv{Y},\bv{\omega}) =
\frac{\Gamma\left(\sum_{i=1}^N\omega_i\right)}
{\prod_{i=1}^N\Gamma(\omega_i)}\prod_{i=1}^N
Y_i^{\omega_i-1},\label{eq:D}
\end{equation}
with $\omega_i=\alpha_i$, $i=1,\dots,K=N-1$, $\omega_N=\beta_K$, and
$Y_N=1-\sum_{j=1}^KY_j$. In the univariate case, $K=N-1=1$,
$\bv{Y}=(Y_1,Y_2)=(Y,1\!-\!Y)$, both $\mathscr{G}$ and $\mathscr{D}$ yield
the beta distribution
\begin{equation}
\mathscr{B}(Y,\alpha,\beta) =
  \frac{\Gamma(\alpha+\beta)}{\Gamma(\alpha)\Gamma(\beta)}
  Y^{\alpha-1}(1-Y)^{\beta-1},
\end{equation}
with $\omega_1=\alpha$ and $\omega_2=\beta$.

$\mathscr{G}$, $\mathscr{D}$, and $\mathscr{B}$ are zero outside the
$K$-dimensional generalized triangle; the sample spaces are bounded. Compared to
$\mathscr{D}$, there are $K-1$ additional parameters in $\mathscr{G}$ for a set
of $K$ scalars.

\subsection{Moments}
All moments of the generalized Dirichlet distribution, \Eqre{eq:GD}, can be
obtained from $\alpha_i$ and $\beta_i$ of which the first two are
\cite{Connor_69,Wong_98}
{\allowdisplaybreaks
\begin{align}
\irmean{Y_i} & = \int Y_i\mathscr{G}(\bv{Y})\mathrm{d}\bv{Y} =
\frac{\alpha_i}{\alpha_i+\beta_i} \prod_{j=1}^{i-1}
\frac{\beta_j}{\alpha_j+\beta_j},\label{eq:GDmeans}\\
\irmean{y_iy_j} & = \irmean{(Y_i-\rmean{Y_i})(Y_j-\rmean{Y_j})} = \left\{
\begin{array}{lr}
\displaystyle\irmean{Y_i}\left(\frac{\alpha_i+1}{\alpha_i+\beta_i+1} M_{i-1} -
\irmean{Y_i}\right) &\qquad \mathrm{for} \quad i = j,\vspace{6pt}\\
\displaystyle\irmean{Y_j}\left(\frac{\alpha_i}{\alpha_i+\beta_i+1} M_{i-1} -
\irmean{Y_i}\right) & \qquad \mathrm{for} \quad i \ne j,
\end{array}
\right.\label{eq:GDcovariances}\\
& \qquad\qquad\qquad\qquad\qquad\qquad\qquad i,j = 1,\dots,K,\nonumber
\end{align}}%
where $M_{i-1} = \prod_{k=1}^{i-1} (\beta_k+1)/(\alpha_k+\beta_k+1)$. Setting
$\gamma_1=\dots=\gamma_{K-1}=0$, with $\omega_i=\alpha_i$, $i=1,\dots,K=N-1$,
$\omega_N=\beta_K$, in Eqs.\ (\ref{eq:GDmeans}--\ref{eq:GDcovariances}) reduces
to the first two moments of the Dirichlet distribution,
\begin{align}
\irmean{Y_i} & = \frac{\omega_i}{\omega},\label{eq:Dmeans}\\
\irmean{y_i y_j} & = \left\{
\begin{array}{lr}
\displaystyle\frac{\omega_i(\omega-\omega_i)}{\omega^2(\omega+1)} &
\quad \mathrm{for} \quad i = j,\vspace{6pt}\\
\displaystyle\frac{-\omega_i \omega_j}{\omega^2(\omega+1)} & \quad
\mathrm{for} \quad i \ne j,
\end{array}\label{eq:Dcovariances}
\right.\\
& \qquad\qquad\qquad i,j = 1,\dots,K,\nonumber
\end{align}
where $\omega\!=\!\sum_{j=1}^N\omega_j$. \Eqre{eq:GDcovariances} shows that in
the generalized Dirichlet distribution $Y_1$ is always negatively correlated
with the other scalars. However, $Y_j$ and $Y_m$ can be positively correlated
for $j,m>1$, see also \cite{Lochner_75}.  According to Wong \cite{Wong_98},
\emph{``If there exists some $m>j$ such that $Y_j$ and $Y_m$ are positively
(negatively) correlated, then $Y_j$ will be positively (negatively) correlated
with $Y_n$ for all $n>j$.''} This can be seen from \Eqre{eq:GDcovariances}: the
sign of $\irmean{y_my_j}$ is independent of $j$, so the sign of
$\irmean{y_my_j}$, $m>j$ will imply the signs of all $\irmean{y_ny_j}$, $n>j$.
This is in contrast with the Dirichlet distribution, \Eqre{eq:D}, whose
covariances are always non-positive as can be seen from \Eqre{eq:Dcovariances}.

In the univariate case, $K=N-1=1$, $\bv{Y}=(Y_1,Y_2)=(Y,1\!-\!Y)$, the first two
moments of both the generalized and the standard Dirichlet distributions, Eqs.\
(\ref{eq:GDmeans}--\ref{eq:GDcovariances}) and Eqs.\
(\ref{eq:Dmeans}--\ref{eq:Dcovariances}), respectively, reduce to the moments
of the beta distribution, with $\omega_1=\alpha$ and $\omega_2=\beta$,
\begin{align}
\irmean{Y} & = \frac{\alpha}{\alpha+\beta},\\
\irmean{y^2} & = \frac{\alpha \beta}{(\alpha+\beta)^2(\alpha+\beta+1)}.
\end{align}

\section{Relation to other diffusion processes}
It also useful to relate the generalized Dirichlet process, \Eqre{eq:iSDE}, to
other multivariate diffusion processes with linear drift and quadratic
diffusion.

Setting $c_{1i}/\kappa_i=\dots=c_{ii}/\kappa_i=1$ for
$i=1,\dots,K\!-\!1$, in \Eqre{eq:iSDE} yields
\begin{equation}
\mathrm{d}Y_i(t) = \frac{b_i}{2} \big[S_i Y_N - (1-S_i)Y_i\big] \mathrm{d}t +
\sqrt{\kappa_i Y_i Y_N} \mathrm{d}W_i(t), \qquad i=1,\dots,K=N-1,
\label{eq:DSDE}
\end{equation}
with $Y_N\!=\!1\!-\!\sum_{j=1}^{N-1}Y_j$ whose invariant is the (standard)
Dirichlet distribution, \Eqre{eq:D}. \Eqre{eq:DSDE} is discussed in
\cite{Bakosi_dir}. Another diffusion process whose invariant is also Dirichlet
is the multivariate Wright-Fisher process \cite{Steinrucken_2013},
\begin{equation}
\mathrm{d}Y_i(t) = \frac{1}{2} (\omega_i-\omega Y_i) \mathrm{d}t
+ \sum_{j=1}^{K} \sqrt{Y_i(\delta_{ij}-Y_j)}
\mathrm{d}W_{ij}(t), \qquad i = 1,\dots,K=N-1,
\label{eq:WF}
\end{equation}
where $\delta_{ij}$ is Kronecker's delta. Another process similar to
\Eqres{eq:iSDE}, \Eqr{eq:DSDE}, and \Eqr{eq:WF} is the multivariate Jacobi
process, used in econometrics,
\begin{equation}
\mathrm{d}Y_i(t) = a(Y_i-\pi_i)\mathrm{d}t + \sqrt{cY_i}\mathrm{d}W_i(t) -
\sum_{j=1}^{N-1} Y_i\sqrt{cY_j}\mathrm{d}W_j(t), \qquad i = 1,\dots,N
\label{eq:Jacobi}
\end{equation}
of Gourieroux \& Jasiak \cite{Gourieroux_06} with $a<0$, $c>0$, $\pi_\alpha>0$,
and $\sum_{j=1}^N\pi_j=1$.

In the univariate case, $K=N-1=1$, $\bv{Y}=(Y_1,Y_2)=(Y,1\!-\!Y)$, the
generalized Dirichlet, Dirichlet, Wright-Fisher, and Jacobi diffusions,
\Eqres{eq:iSDE}, \Eqr{eq:DSDE}, \Eqr{eq:WF}, \Eqr{eq:Jacobi}, respectively, all
reduce to
\begin{equation}
\mathrm{d}Y(t) = \frac{b}{2} (S-Y)\mathrm{d}t +
                 \sqrt{\kappa Y(1-Y)}\mathrm{d}W(t),
\label{eq:beta}
\end{equation}
see also \cite{Bakosi_beta}, whose invariant is the beta distribution, which
belongs to the family of Pearson diffusions, discussed in detail by Forman \&
Sorensen \cite{Forman_08}.

\section{Summary}
\label{sec:summary}
Following the development in \cite{Bakosi_dir} we started with a multivariate
distribution for a set of stochastic variables that satisfies a conservation
principle in which all variables sum to unity. Applying the constraints on the
existence of potential solutions of Fokker-Planck equations, we derived a system
of stochastic differential equations \Eqr{eq:iSDE} whose joint distribution in
the statistically stationary state is Lochner's generalized Dirichlet
distribution, \Eqre{eq:GD}. \Eqre{eq:iSDE} is a generalization of the Dirichlet
diffusion process developed in \cite{Bakosi_dir}. Compared to the standard
Dirichlet process, the generalized diffusion allows for representing a more
general class of stochastic processes with a more general covariance matrix. The
process may be stationary or non-stationary, not limited to non-positive
covariances, and satisfies the unit-sum requirement, \Eqre{eq:sum}, at all
times, necessary for variables that obey a conservation principle.

\bibliographystyle{unsrt}
\bibliography{jbakosi}

\providecommand{\noopsort}[1]{}
\begin{thebibliography}{10}

\bibitem{Lochner_75}
R.~H. Lochner.
\newblock A {G}eneralized {D}irichlet {D}istribution in {B}ayesian {L}ife
  {T}esting.
\newblock {\em Journal of the Royal Statistical Society. Series B
  (Methodological)}, 37(1):pp. 103--113, 1975.

\bibitem{Bakosi_dir}
J.~Bakosi and J.R. Ristorcelli.
\newblock A stochastic diffusion process for the {D}irichlet distribution.
\newblock {\em Int. J. Stoch. Anal.}, 2013:7, 2013.
\newblock Article ID 842981.

\bibitem{Connor_69}
R.~J. Connor and J.~E. Mosimann.
\newblock Concepts of {I}ndependence for {P}roportions with a {G}eneralization
  of the {D}irichlet {D}istribution.
\newblock {\em J. Am. Stat. Assoc.}, 64(325):194--206, 1969.

\bibitem{Wong_98}
Tzu-Tsung Wong.
\newblock Generalized {D}irichlet distribution in {B}ayesian analysis.
\newblock {\em Appl. Math. Comput.}, 97(2-3):165 -- 181, 1998.

\bibitem{Johnson_60}
N.~L. Johnson.
\newblock An approximation to the multinomial distribution some properties and
  applications.
\newblock {\em Biometrika}, 47(1-2):93--102, 1960.

\bibitem{Mosimann_62}
J.~E. Mosimann.
\newblock On the compound multinomial distribution, the
  multivariate-distribution, and correlations among proportions.
\newblock {\em Biometrika}, 49(1-2):65--82, 1962.

\bibitem{Kotz_00}
S.~Kotz, N.L. Johnson, and N.~Balakrishnan.
\newblock {\em {Continuous Multivariate Distributions: Models and
  applications}}.
\newblock Wiley series in probability and statistics: Applied probability and
  statistics. Wiley, 2000.

\bibitem{Pearson_1896}
K.~Pearson.
\newblock {Mathematical Contributions to the Theory of Evolution. On a Form of
  Spurious Correlation Which May Arise When Indices Are Used in the Measurement
  of Organs}.
\newblock {\em Royal Society of London Proceedings Series I}, 60:489--498,
  1896.

\bibitem{Paulino_95}
C.D.M. Paulino and de~Braganca Pereira~C.A.
\newblock Bayesian methods for categorical data under informative general
  censoring.
\newblock {\em Biometrika}, 82(2):439--446, 1995.

\bibitem{Chayes_62}
F.~Chayes.
\newblock Numerical correlation and petrographic variation.
\newblock {\em J. Geol.}, 70:440--452, 1962.

\bibitem{Martin_65}
P.~S. Martin and J.~E. Mosimann.
\newblock Geochronology of pluvial l{a}ke {C}ochise, southern {A}rizona; [part]
  3, {P}ollen statistics and {P}leistocene metastability.
\newblock {\em Am. J. Sci.}, 263:313--358, 1965.

\bibitem{Lange_95}
K.~Lange.
\newblock Applications of the {D}irichlet distribution to forensic match
  probabilities.
\newblock {\em Genetica}, 96:107--117, 1995.
\newblock 10.1007/BF01441156.

\bibitem{Gourieroux_06}
C.~Gourieroux and J.~Jasiak.
\newblock {M}ultivariate {J}acobi process with application to smooth
  transitions.
\newblock {\em Journal of Econometrics}, 131:475--505, 2006.

\bibitem{Girimaji_91}
S.~S. Girimaji.
\newblock Assumed $\beta$-pdf model for turbulent mixing: validation and
  extension to multiple scalar mixing.
\newblock {\em Combust. Sci. Technol.}, 78(4):177 -- 196, 1991.

\bibitem{Steinrucken_2013}
M.~Steinrucken, Y.X.~Rachel Wang, and Y.S. Song.
\newblock An explicit transition density expansion for a multi-allelic
  {W}right--{F}isher diffusion with general diploid selection.
\newblock {\em Theoretical Population Biology}, 83(0):1--14, 2013.

\bibitem{Gardiner_09}
C.~W. Gardiner.
\newblock {\em Sto\-chastic methods, A Handbook for the Natural and Social
  Sciences}.
\newblock Springer-Verlag, Berlin Heidelberg, 4 edition, 2009.

\bibitem{Bakosi_beta}
J.~Bakosi and J.R. Ristorcelli.
\newblock Exploring the beta distribution in variable-density turbulent mixing.
\newblock {\em J. Turbul.}, 11(37):1--31, 2010.

\bibitem{Forman_08}
J.~L. Forman and M.~Sorensen.
\newblock The {P}earson {D}iffusions: {A} {C}lass of {S}tatistically
  {T}ractable {D}iffusion {P}rocesses.
\newblock {\em Scandinavian Journal of Statistics}, 35:438--465, 2008.

\bibitem{Kloeden_99}
P.~E. Kloeden and E.~Platen.
\newblock {\em {N}umerical {S}olution of {S}tochastic {D}ifferential
  {E}quations}.
\newblock Springer, Berlin, 1999.

\end{thebibliography}

\newpage

\section*{Appendix A: Inductive proof of Eqs.\
(\ref{eq:galphai}--\ref{eq:gammaN}) based on $\bv{K=3}$}
Eqs.\ (\ref{eq:galphai}--\ref{eq:gammaN}) are now arrived at for $K=3$, yielding
the correspondence of the generalized Dirichlet distribution, \Eqre{eq:GD}, and
its stochastic process, \Eqre{eq:iSDE}, for $K=3$. The procedure generalizes to
arbitrary $K>3$.

From \Eqre{eq:gphi} the scalar potential for $K=3$ is
\begin{align}
-\phi(Y_1,Y_2,Y_3) & = (\alpha_1-1)\ln Y_1 +
                       (\alpha_2-1)\ln Y_2 +
                       (\alpha_3-1)\ln Y_3 \nonumber \\
             & \quad + \gamma_1\ln(1-Y_1) +
                       \gamma_2\ln(1-Y_1-Y_2) +
                       \gamma_3\ln(1-Y_1-Y_2-Y_3).
\label{eq:gphi3}
\end{align}
From Eqs.\ (\ref{eq:ga}--\ref{eq:gB}) the drift and diffusion for $K=3$ are
\begin{align}
a_1 & = \frac{b_1/2}{(1-Y_1)(1-Y_1-Y_2)}
        \Big[S_1(1-Y_1-Y_2-Y_3) - (1-S_1)Y_1\Big] \nonumber \\
    & \qquad + \frac{Y_1(1-Y_1-Y_2-Y_3)}{(1-Y_1)(1-Y_1-Y_2)}
        \left[\frac{c_{11}/2}{1-Y_1} + \frac{c_{12}/2}{1-Y_1-Y_2} \right],
\\[0.3cm]
a_2 & = \frac{b_2/2}{1-Y_1-Y_2}
        \Big[S_2(1-Y_1-Y_2-Y_3) - (1-S_2)Y_2\Big]
      + \frac{c_{22}}{2}\!\cdot\!\frac{Y_2(1-Y_1-Y_2-Y_3)}{(1-Y_1-Y_2)^2},
\\[0.3cm]
a_3 & = \frac{b_3}{2}\Big[S_3(1-Y_1-Y_2-Y_3) - (1-S_3)Y_3\Big], \\[0.3cm]
B_{11} & = \kappa_1 \frac{Y_1(1-Y_1-Y_2-Y_3)}{(1-Y_1)(1-Y_1-Y_2)}, \\[0.3cm]
B_{22} & = \kappa_2 \frac{Y_2(1-Y_1-Y_2-Y_3)}{1-Y_1-Y_2}, \\[0.3cm]
B_{33} & = \kappa_3 Y_3 (1-Y_1-Y_2-Y_3), \\[0.3cm]
B_{12} & = B_{23} = B_{13} = 0, \label{eq:offDiagDiff}
\end{align}
Substituting Eqs.\ (\ref{eq:gphi3}--\ref{eq:offDiagDiff}) into
\Eqre{eq:solution} for $K\!=\!3$ yields
\begin{align}
\frac{\alpha_1-1}{Y_1}
- \frac{\gamma_1}{1-Y_1}
- \frac{\gamma_2}{1-Y_1-Y_2}
- \frac{\gamma_3}{1-Y_1-Y_2-Y_3} = \nonumber 
\qquad\qquad\qquad\qquad\qquad\qquad\qquad \\
= \left(\frac{b_1}{\kappa_1}S_1-1\right) \frac{1}{Y_1}
+ \left(\frac{c_{11}}{\kappa_1}-1\right) \frac{1}{1-Y_1}
+ \left(\frac{c_{12}}{\kappa_1}-1\right) \frac{1}{1-Y_1-Y_2} \nonumber \\
+ \left[1-\frac{b_1}{\kappa_1}(1-S_1) \right] \frac{1}{1-Y_1-Y_2-Y_3},
\label{eq:corr1}\\[0.3cm]
\frac{\alpha_2-1}{Y_2}
- \frac{\gamma_2}{1-Y_1-Y_2}
- \frac{\gamma_3}{1-Y_1-Y_2-Y_3} = \nonumber
\qquad\qquad\qquad\qquad\qquad\qquad\qquad\qquad\qquad \\
= \left(\frac{b_2}{\kappa_2}S_2-1\right) \frac{1}{Y_2}
+ \left(\frac{c_{22}}{\kappa_2}-1\right) \frac{1}{1-Y_1-Y_2}
+ \left[1-\frac{b_2}{\kappa_2}(1-S_2) \right] \frac{1}{1-Y_1-Y_2-Y_3},
\label{eq:corr2}\\[0.3cm]
\frac{\alpha_3-1}{Y_3}
- \frac{\gamma_3}{1-Y_1-Y_2-Y_3} =
  \left(\frac{b_3}{\kappa_3}S_3-1\right) \frac{1}{Y_3}
+ \left[1-\frac{b_3}{\kappa_3}(1-S_3) \right] \frac{1}{1-Y_1-Y_2-Y_3},
\end{align}
which shows that if
\begin{align}
\alpha_1 & = \frac{b_1}{\kappa_1}S_1, \label{eq:a1} \\
\alpha_2 & = \frac{b_2}{\kappa_2}S_2, \\
\alpha_3 & = \frac{b_3}{\kappa_3}S_3, \\
1-\gamma_1 & = \frac{c_{11}}{\kappa_1}, \\
1-\gamma_2 & = \frac{c_{12}}{\kappa_1} = \frac{c_{22}}{\kappa_2}, \\
1+\gamma_3 & = \frac{b_1}{\kappa_1}(1-S_1)
             = \frac{b_2}{\kappa_2}(1-S_2)
             = \frac{b_3}{\kappa_3}(1-S_3), \label{eq:g3}
\end{align}
all hold, the invariant of \Eqre{eq:iSDE} is \Eqre{eq:GD} for $K=3$,
\begin{align}
\mathscr{G}(Y_1,Y_2,Y_3,\alpha_1,\alpha_2,\alpha_3,\beta_1,\beta_2,\beta_3) =
\qquad &\nonumber \\
\frac{\Gamma(\alpha_1+\beta_1) \Gamma(\alpha_2+\beta_2)\Gamma(\alpha_3+\beta_3)}
     {\Gamma(\alpha_1)\Gamma(\beta_1)
      \Gamma(\alpha_2)\Gamma(\beta_2)
      \Gamma(\alpha_3)\Gamma(\beta_3)}
\label{eq:GD3}
\times &\\
\times Y_1^{\alpha_1-1} Y_2^{\alpha_2-1} Y_3^{\alpha_3-1}
(1-Y_1)^{\gamma_1} & (1-Y_1-Y_2)^{\gamma_2} (1-Y_1-Y_2-Y_3)^{\gamma_3},\nonumber
\end{align}
with
\begin{equation}
\gamma_1 = \beta_1 - \alpha_2 - \beta_2, \qquad
\gamma_2 = \beta_2 - \alpha_3 - \beta_3, \qquad
\gamma_3 = \beta_3 - 1. \label{eq:g}
\end{equation}
Eqs.\ (\ref{eq:a1}--\ref{eq:g3}) give the correspondence between the
coefficients of the stochastic system, \Eqre{eq:iSDE}, and its invariant,
\Eqre{eq:GD}, for $K=3$. With \Eqre{eq:g} the correspondence between the
parameters of the joint probability density function (PDF),
($\alpha_1,\alpha_2,\alpha_3,\beta_1,\beta_2,\beta_3$), and the coefficients of
the stochastic system,
($b_1,b_2,b_3,$ $S_1,S_2,S_3,$ $\kappa_1,\kappa_2,\kappa_3,c_{11},c_{12},c_{22}$), is
also given.

It is straightforward to verify that setting $\gamma_1 = \gamma_2 = 0$ in
\Eqre{eq:GD3} yields the Dirichlet distribution, \Eqre{eq:D}, for $K=3$
~($N=4)$,
\begin{align}
\mathscr{D}(Y_1,Y_2,Y_3,\omega_1,\omega_2,\omega_3,\omega_4) =
&\nonumber \\
\frac{\Gamma(\omega_1+\omega_2+\omega_3+\omega_4)}
     {\Gamma(\omega_1)\Gamma(\omega_2)\Gamma(\omega_3)\Gamma(\omega_4)} &
Y_1^{\omega_1-1} Y_2^{\omega_2-1} Y_3^{\omega_3-1} (1-Y_1-Y_2-Y_3)^{\omega_4-1}
\end{align}
with
\begin{equation}
\omega_1 = \alpha_1, \qquad
\omega_2 = \alpha_2, \qquad
\omega_3 = \alpha_3, \qquad
\omega_4 = \beta_3.
\end{equation}
Similarly, setting $c_{11}/\kappa_1 = c_{12}/\kappa_1 =  c_{22}/\kappa_2 = 1$
in Eqs.\ (\ref{eq:corr1}--\ref{eq:corr2}) reduces to the system corresponding
that of the Dirichlet case \cite{Bakosi_dir}.

\section*{Appendix B: Numerical simulation: The effect of the extra coefficient
for $\bv{K=2}$}%
Numerical simulations are used to demonstrate the effect of the extra
coefficient, $c_{11}$, compared to the standard Dirichlet case, given in
\cite{Bakosi_dir}.

The time-evolution of an ensemble of $10,\!000$ particles has been numerically
computed by integrating the system \Eqr{eq:Ito}, with drift and
diffusion (\ref{eq:ga}--\ref{eq:gB}), for $K=2$, i.e., ($Y_1, Y_2, Y_3 = 1 - Y_1
- Y_2$),
\begin{align}
\mathrm{d}Y_1 & = \frac{b_1/2}{1-Y_1} \big[S_1 Y_3 - (1-S_1)Y_1\big]\mathrm{d}t
+ \frac{Y_1Y_3}{1-Y_1} \cdot \frac{c_{11}/2}{1-Y_1} \mathrm{d}t + \sqrt{\kappa_1
\frac{Y_1 Y_3}{1-Y_1}}\mathrm{d}W_1, \label{eq:Ito1}\\
\mathrm{d}Y_2 & = \frac{b_2}{2}\big[S_2 Y_3 - (1-S_2)Y_2\big]\mathrm{d}t +
\sqrt{\kappa_2 Y_2 Y_3}\mathrm{d}W_2, \label{eq:Ito2}\\
Y_3 & = 1 - Y_1 - Y_2. \label{eq:Ito3}
\end{align}%
\begin{figure}[!]
  \makeatletter
  \renewcommand{\fnum@figure}{Tab.~\thefigure}
  \makeatother
  \caption{\label{tab:coeff}Coefficients of Eqs.\ (\ref{eq:Ito1}--\ref{eq:Ito3})
           and asymptotic moments for three simulation cases.}
  \centering
  \noindent\rule{\textwidth}{0.4pt}
  \begin{subfigure}[t]{0.51\textwidth}
     Asymptotic moments for $K=2$, \\ see Eqs.\
     (\ref{eq:GDmeans}--\ref{eq:GDcovariances})
     \begin{align*}
       \rmean{Y_1} &= \frac{\alpha_1}{\alpha_1 + \beta_1} \\
       \rmean{Y_2} &= \frac{\alpha_2}{\alpha_2 + \beta_2} \cdot
       \frac{\alpha_1}{\alpha_1 + \beta_1} \\
       \rv{y_1} &= \rmean{Y_1}\left( \frac{\alpha_1 + 1}{\alpha_1 + \beta_1 + 1}
       - \rmean{Y_1} \right) \\
       \rv{y_2} &= \rmean{Y_2}\left( \frac{\alpha_2 + 1}{\alpha_2 + \beta_2 + 1}
       \cdot \frac{\alpha_1 + 1}{\alpha_1 + \beta_1 + 1} - \rmean{Y_2} \right)\\
       \rmean{y_1y_2} &= \rmean{Y_2}\left( \frac{\alpha_1}{\alpha_1 + \beta_1 +
       1} - \rmean{Y_1} \right)
     \end{align*}
  \end{subfigure}
  \begin{subfigure}[t]{0.48\textwidth}
     Dirichlet SDE coefficients (common to all cases)
     \begin{equation*}
       \begin{aligned}[c]
               b_1 &= 1/10 \\
               S_1 &= 5/8  \\
          \kappa_1 &= 1/80 \\
       \end{aligned} \qquad
       \begin{aligned}[c]
              b_2 &= 3/2 \\
              S_2 &= 2/5 \\
         \kappa_2 &= 3/10
       \end{aligned}
     \end{equation*} \\[0.3cm]
     Generalized Dirichlet SDE coefficients
     \begin{equation*}
       \begin{aligned}[c]
           c_{11} &= \kappa_{11} = 1/80  \\
           c_{11} &= -1/80 \\
           c_{11} &= -1/4  \\
       \end{aligned} \quad
       \begin{aligned}[c]
           \textrm{(case 1)} \\
           \textrm{(case 2)} \\
           \textrm{(case 3)}
       \end{aligned}
     \end{equation*}
  \end{subfigure}
  \\[0.5cm]
  \begin{subfigure}[t]{0.5\textwidth}
     PDF parameters from the SDE coefficients, \\ see Eqs.\
     (\ref{eq:galphai}--\ref{eq:gammaN})
     \begin{align*}
         \alpha_1 &= \frac{b_1}{\kappa_1}S_1 \\
         \alpha_2 &= \frac{b_2}{\kappa_2}S_2 \\
       1-\gamma_1 &= \frac{c_{11}}{\kappa_1} \\
       1+\gamma_2 &= \frac{b_1}{\kappa_1}(1-S_1) = \frac{b_2}{\kappa_2}(1-S_2)\\
          \beta_2 &= 1 + \gamma_2 = \frac{b_1}{\kappa_1}(1-S_1) =
                     \frac{b_2}{\kappa_2}(1-S_2) \\
          \beta_1 &= \gamma_1 + \alpha_2 + \beta_2 = 1 - \frac{c_{11}}{\kappa_1}
                     + \alpha_2 + \beta_2 \\
     \end{align*}
  \end{subfigure}
  \begin{subfigure}[t]{0.49\textwidth}
     SDE asymptotic moments for cases 1, 2, 3
     \begin{equation*}
       \begin{aligned}
                 c_{11} &= \frac{1}{80}         \\
               \alpha_1 &= 5                    \\
               \alpha_2 &= 2                    \\
                \beta_2 &= 3                    \\
                \beta_1 &= 5                    \\
            \rmean{Y_1} &= \frac{1}{2}          \\
            \rmean{Y_2} &= \frac{1}{5}          \\
               \rv{y_1} &= \frac{1}{44}         \\
               \rv{y_2} &= \frac{4}{275}        \\
         \rmean{y_1y_2} &= -\frac{1}{110}
       \end{aligned} \quad
       \begin{aligned}
                 c_{11} &= -\frac{1}{80}        \\
               \alpha_1 &= 5                    \\
               \alpha_2 &= 2                    \\
                \beta_2 &= 3                    \\
                \beta_1 &= 7                    \\
            \rmean{Y_1} &= \frac{5}{12}         \\
            \rmean{Y_2} &= \frac{7}{30}         \\
               \rv{y_1} &= \frac{35}{1872}      \\
               \rv{y_2} &= \frac{609}{35100}    \\
         \rmean{y_1y_2} &= -\frac{35}{4680}
       \end{aligned} \quad
       \begin{aligned}
                 c_{11} &= -\frac{1}{4}         \\
               \alpha_1 &= 5                    \\
               \alpha_2 &= 2                    \\
                \beta_2 &= 3                    \\
                \beta_1 &= 26                   \\
            \rmean{Y_1} &= \frac{5}{31}         \\
            \rmean{Y_2} &= \frac{52}{155}       \\
               \rv{y_1} &= \frac{65}{15376}     \\
               \rv{y_2} &= \frac{11141}{384400} \\
         \rmean{y_1y_2} &= -\frac{13}{7688}
       \end{aligned}
     \end{equation*}
  \end{subfigure}\\[0.3cm]
  \noindent\rule{\textwidth}{0.4pt}
\end{figure}
\setcounter{figure}{0}%
In Eqs.\ (\ref{eq:Ito1}--\ref{eq:Ito2}) $\mathrm{d}W_1$ and $\mathrm{d}W_2$ are
independent Wiener processes, sampled from Gaussian streams of random numbers
with mean $\irmean{\mathrm{d}W_i}\!=\!0$ and covariance $\irmean{\mathrm{d} W_i
\mathrm{d} W_j}\!= \!\delta_{ij} \mathrm{d}t$. Eqs.\
(\ref{eq:Ito1}--\ref{eq:Ito3}) were advanced in time with the Euler-Maruyama
scheme \cite{Kloeden_99} with time step $\Delta t\!=\!0.025$. The coefficients
of the stochastic system (\ref{eq:Ito1}--\ref{eq:Ito3}), the corresponding
parameters and the first two moments of the asymptotic generalized Dirichlet
distributions for $K\!=\!2$ are shown in Table \ref{tab:coeff}. Three different
cases were simulated. Here the initial condition of $(Y_1,Y_2) \equiv 0$ was
used.  The initial PDF in all cases is the same: all samples are zero and the
PDF is therefore not Dirichlet nor Generalized Dirichlet, see also
\cite{Bakosi_dir} for nonzero but different non-Dirichlet initial conditions.
Our motivation is two-fold: (1) to show that the solution approaches the
invariant, and (2) to show how the new additional parameter in the generalized
Dirichlet SDE affects the dynamics. Had the initial conditions coincided with
the given invariant, the PDF (and its statistics) would not have changed in time
-- as has been demonstrated mathematically. The SDE coefficients in the three
simulations only differ in the extra generalized Dirichlet coefficient,
$c_{11}$, otherwise, the setup corresponds to the example in \cite{Bakosi_dir}.
In the first simulation $c_{11}\!=\kappa_1\!=\!1/80$, i.e., $c_{11}$ is
\emph{not} a free coefficient and is chosen to yield an asymptotic solution that
is a (standard) Dirichlet, the same as in \cite{Bakosi_dir}. In the second and
third simulations $c_{11}$ are freely chosen and thus yield generalized
Dirichlet solutions.  Figure~\ref{fig:sim} shows the evolutions of the first two
moments in time for the three cases.
\begin{figure}[!]
\centering
\resizebox{1.0\columnwidth}{!}{\input{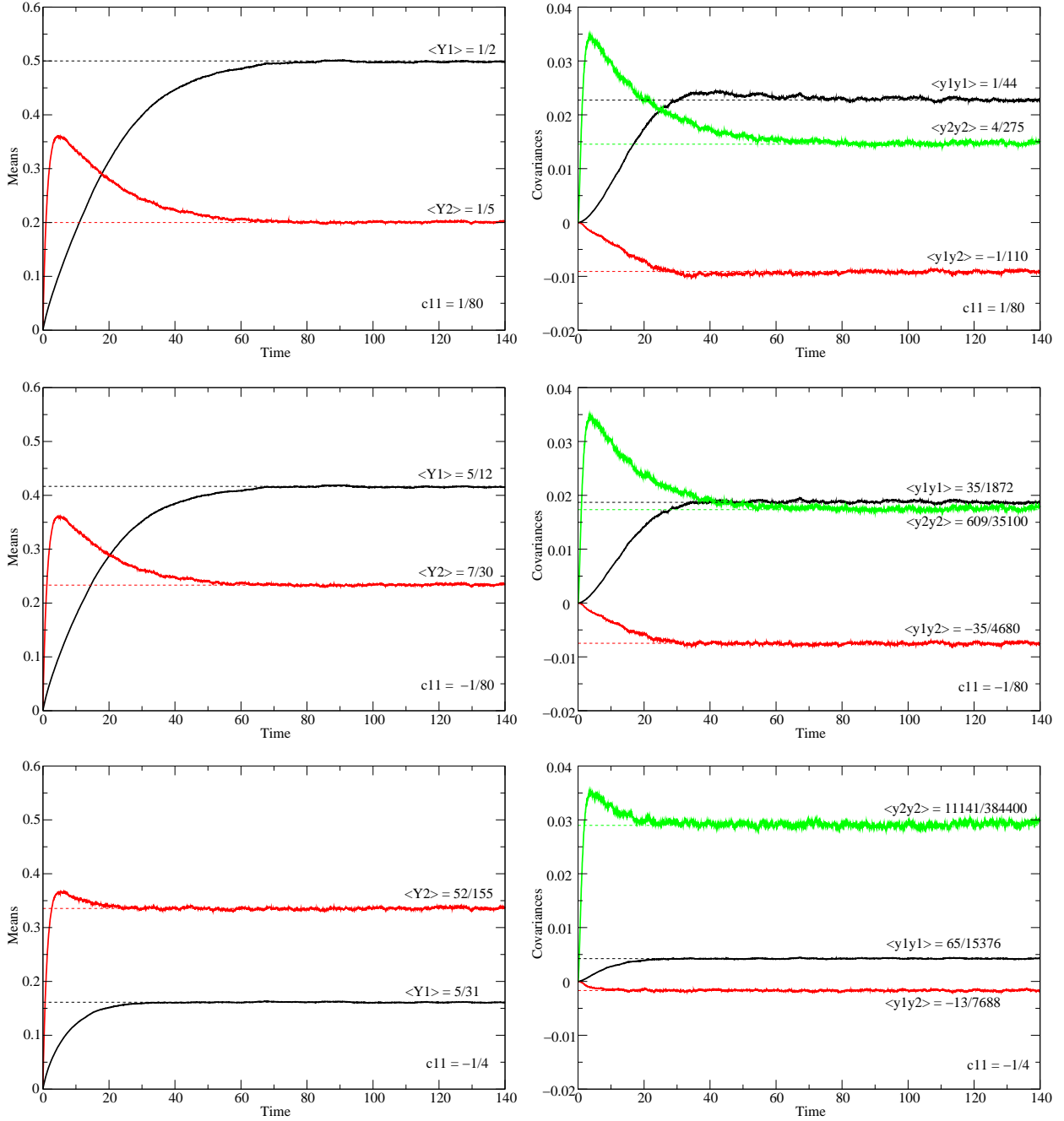}}
\caption{Time evolution of the first two moments of Eqs.\
(\ref{eq:Ito1}--\ref{eq:Ito3}). First row: $c_{11}\!=\!\kappa_1\!=\!1/80$
(standard Dirichlet, see also \cite{Bakosi_dir}), second row:
$c_{11}\!=\!-1/80$, third row: $c_{11}\!=\!-1/4$.}
\label{fig:sim}
\end{figure}

\end{document}